\documentclass[12pt,preprint]{aastex}
\usepackage{times}
\usepackage{natbib}
\newcommand{\rsun}{\mbox{${\rm R}_{\odot}$}}
\newcommand{\msun}{\mbox{${\rm M}_{\odot}$}}

\newcommand{\simgt}{\lower.5ex\hbox{$\; \buildrel > \over \sim \;$}}
\newcommand{\simlt}{\lower.5ex\hbox{$\; \buildrel < \over \sim \;$}}
\newcommand{\BV}{Brunt-V\"ais\"al\"a\ }
\def\apj{ApJ}%
\def\apjl{ApJ}%
\def\apjs{ApJS}%
\def\ao{Appl.~Opt.}%
\def\apss{Ap\&SS}%
\def\aap{A\&A}%
\def\mnras{MNRAS}%
\def\solphys{Sol.~Phys.}%
\def\nat{Nature}%
\slugcomment{Not to appear in Nonlearned J., 45.}

\shorttitle{short title}
\shortauthors{Montalb\'{a}n et al.}

\begin{document}


\title{Testing convective-core overshooting using period spacings of dipole modes in red giants}


\author{J. Montalb\'an}
\affil{Institut d'Astrophysique et G\'eophysique de l'Universit\'e de Li\`ege, All\'ee du six Ao\^ut, 17 B-4000 Li\`ege, Belgium}

\author{A. Miglio }
\affil{School of Physics and Astronomy, University of Birmingham, Edgbaston, Birmingham B15 2TT, United Kingdom}

\author{A. Noels, M.-A, Dupret, R. Scuflaire}
\affil{Institut d'Astrophysique et G\'eophysique de l'Universit\'e de Li\`ege, All\'ee du six Ao\^ut, 17 B-4000 Li\`ege, Belgium}

\and

\author{P. Ventura}
\affil{Osservatorio Astronomico di Roma-INAF, via Frascati 33, Monteporzio Catone, Rome, Italy}



\begin{abstract}
Uncertainties on central mixing in main sequence (MS) and core He-burning (He-B)  phases affect key predictions of stellar evolution such as late evolutionary
 phases, chemical enrichment, ages etc.  We propose a test of the extension of extra-mixing in two relevant evolutionary phases based on
period spacing ($\Delta P$) of solar-like oscillating giants.  From stellar models and
their corresponding adiabatic frequencies (respectively computed with
ATON and LOSC codes) we provide the first predictions of the
observable $\Delta P$  for stars in the red giant branch (RGB) and in
the red clump (RC). We find:  {\it i)} a clear correlation between
$\Delta P$ and the mass of the helium core ($M_{\rm He}$); the latter
in intermediate-mass stars depends on the MS overshooting, hence it
can be used to set constraints on extra mixing during MS when coupled
with chemical composition; {\it ii)}  a linear dependence of the 
average value of  the asymptotic period spacing  ($\langle \Delta P \rangle_a$) during the He-B phase  on the size of the
convective core.  A first comparison with the inferred asymptotic period
spacing for {\it Kepler} RC stars suggests the need for extra mixing also during this phase, as evinced from other observational facts.
\end{abstract}


\keywords{Stars: evolution --- stars: interiors --- stars: oscillations --- stars: late-type}



\section{Introduction}

Despite the numerous efforts undertaken to understand convection in stars, the treatment of this process in stellar modelling is still rather simplistic and  one of the major uncertainties affecting the predictions of stellar evolution theory. In particular, whilst convection is a highly non-local phenomenon, the extension of the convective region is determined by local criteria such as Schwarzschild and Ledoux ones. Nowadays, there are clear evidences  that the current description of convection is  unsatisfactory: numerical simulations, laboratory experiments, and disagreement between theoretical predictions  and observations.  For instance,  mixing beyond the convective core formal limit  during Main Sequence (MS) is needed to reproduce the morphology of color-magnitude diagrams of stellar cluster and the properties of  binary systems \citep[see e.g.][]{maedermermilliod1981,andersenetal1990,ribasetal2000}. Similarly, observational evidences suggest that the extension of the central mixed region during the core He-burning phase should be larger than determined by the Schwarzschild criterion.  In fact, this  extension has important consequences on the duration of the He-B phase, but also in the following evolutionary phases, determining, for instance, the ratio between asymptotic giant branch (AGB) and horizontal branch (HB) stars. Moreover, the different chemical profiles of C and O from different kind of mixing directly affect the oxygen abundance of white dwarfs \citep[see e.g.][]{stranieroetal2003}. 
The nature of the mechanism(s) inducing extra-mixing, both in  MS and He-B phases,  as well as its extension are still debated \citep[see e.g.][for a review]{chiosi2007}.

Stellar seismology tries to answer some of these questions by looking for seismic indexes based on different oscillation modes and asymptotic  relationships, or deviations with respect to them \citep[see e.g.][for a review]{noelsetal2010}. In this context, the most powerful diagnostics are those based on oscillation modes that propagate close to the central region. In particular gravity modes (g-modes) and mixed gravity-pressure modes. Some of these seismic indexes were successfully applied to individual stars showing different kind of pulsations: solar-like oscillations in MS stars and sub-giants  \citep{dimauroetal2003, migliomontalban2005, deheuvelsmichel2011}  and  B-type pulsators \citep{aertsetal2003,dziembowski2008, degrooteetal2010}. 

Here, we show that solar-like oscillation modes in G-K red giants (RGs) provide a most effective tool to test convective-core overshooting both during the main-sequence and the core-He-burning phase.  Moreover, thanks to  $Kepler$ and CoRoT observations, such tests can potentially be performed on a large number (thousands) of stars encompassing a wide range of stellar parameters.

\section{Red giants: internal structure and evolution}

The evolution of post-MS stars is characterised by a contracting He core and an expanding H-rich envelope, with a H-burning shell in between which becomes thinner and thinner as the star evolves on the RGB. 
The contracting core releases thermal energy, part of which is used to increase the temperature as long as the gas is not too degenerate, but which no longer produces  heating in highly degenerate conditions.  For a given He-core mass   there is a maximum temperature that can be reached by core contraction  as any further contraction leads to  gas cooling. Highly degenerate cores cannot  ignite  Helium burning if their mass is lower than about 0.475~\msun. Such stars keep ascending the RGB until their He cores reach this critical value \citep[see for instance][]{ kippenhahn1990, sweigartetal1990}

For total masses lower than about 2~\msun, the He-core mass at the onset of the He flash and at the start of the post flash He-burning phase (ZAHeB) is thus about the same, i.e. $\sim$0.475~\msun. With the increase of the total mass, the electron degeneracy level is lower and the He-core mass required for He burning decreases. After reaching a minimum value, the He-core mass  increases again following the total mass, as a result of the larger and larger convective core mass during the MS phase \citep[][and references therein]{girardietal1998, castellanietal2000}. Stars with the minimum He-core mass ($\sim 0.33$~\msun, the lowest mass for a pure He star to start He burning--see e.g. Kippenhan and Weigert 1990) define then a transition between two different behaviours of RG evolution \citep{sweigartetal1990}.  Helium-burning stars with masses smaller than the transition stellar mass ($M_{\rm tr}$) populate the so-called  Red Clump \citep[RC,][]{girardietal1998}. Since the luminosity of He-burning stars depends mainly on the He-core mass, the luminosity of RC stars is approximately constant, it rapidly decreases near the transition and then increases drastically as the total mass increases. The stellar mass at which the transition occurs depends on the He-core mass at the end of MS, hence on the chemical composition and on the amount of overshooting  in the MS models. He-burning stars with masses close to the transition mass form the so-called secondary clump \citep{girardi1999}.

The models presented in this paper were computed with the stellar evolution code ATON \citep{venturaetal2008}. We followed the evolution  from pre-main sequence to central He  exhaustion of stellar models with masses from  0.7 to 4.0~\msun,  following the He-flash for low-mass models.   Figure~\ref{fig:modes} (upper panel)  shows the He-core mass at He-ignition as  function of the stellar mass  for  models computed without and with core overshooting during the main sequence evolution. For the  chemical composition considered, the transition mass is 2.4~\msun, while the value decreases to 2.2~\msun\ for models  with overshooting. A similar decrease  is obtained by decreasing the metallicity by a factor of two. 

\section {Adiabatic oscillation properties: period spacing}

The  properties of oscillation modes depend on the behaviour of the \BV\ ($N$) and Lamb ($S_{\ell}$) frequencies. Because of the high density contrast between the core and the envelope,  the RG oscillation spectrum is characterised by  a large number of mixed p-g modes in addition to the radial ones \citep[see for instance][for details]{jcd2004,montalbanetal2010b}. 

In RGs, once the temperature for He ignition is reached,  convection appears in the nuclear burning core and is accompanied by the expansion of the star  central regions. The core structure is, therefore, radically different from that of a RGB star, characterised by a high-density  electron-degenerate radiative helium core (see Fig. \ref{fig:propag}).  Both the presence of a convective core and the decrease of the density contrast ($\rho_{\rm c}/\langle \rho\rangle$) determine significant changes in  the Brunt-V\"ais\"al\"a frequency distribution near the core and, therefore, in the seismic properties of dipole modes \citep{montalbanetal2010b}.   The asymptotic approximation for g-modes  \citep{tassoul1980} predicts that  the periods  of two modes of same degree ($\ell$) and consecutive order ($n$)  are separated by a constant value $\langle\Delta P\rangle_{\rm a}$:

\begin{equation}
\langle \Delta P \rangle_{\rm a} = \frac{2 \pi^2}{\sqrt{\ell (\ell+1)}} \frac{1}{\int N/r dr} 
\label{eq_dp}
\end{equation}

\noindent In the He-burning (He-B) model the expansion of the central layers  leads to a lower $N$  maximum, and its location is displaced at larger stellar radius. Moreover, the central convective regions do not contribute to the integral in Eq.~\ref{eq_dp}  (see Fig. \ref{fig:propag}, upper panel, for a detailed description).  The period spacing between consecutive g-modes, as determined from Eq.~\ref{eq_dp}, is therefore significantly smaller in the RGB model ($80$~s) than in the He-B one ($240$~s).

The asymptotic approximation is, however, no longer valid for mixed modes (those observed in RGs). For a detailed comparisons with the observations we compute therefore adiabatic oscillation frequencies, using the Eulerian version of the code LOSC \citep{scuflaireetal2008}. The results are shown in the lower panels of Fig.~\ref{fig:propag} where we plot the mode inertia of radial and dipole modes, as a function of the frequencies, as well as the period separation  between dipole modes of consecutive radial order $n$ ($\Delta P=P(n+1)-P(n)$).
In addition to a significant difference in the period spacing itself, the differences between the spectra can be summarised as follows: {\it i)} in the RGB model the inertia of $\ell=1$ pressure-dominated modes (corresponding to local minima in $E$) is closer to that of the radial models, indicating a weaker coupling between gravity and acoustic cavities compared to the He-B phase. This is also evident from the gravity-dominated modes of RGB, for which $\Delta P$ is almost constant  (as expected for pure  g-modes) except for the modes describing the minimum of inertia, while for the g-p mixed modes of the He-B model the deviation of $\Delta P$ from a constant value is more important. {\it ii)} The density of $\ell=1$ modes for the He-B model is lower than for the RGB one (by a factor of 3 for the models shown in Fig.~\ref{fig:propag}).

To compare theoretical predictions and observational results it is
mandatory to define theoretical indexes as close as possible to the
observational ones. To derive the average period spacing equivalent to
that measured in \cite{beddingetal2011} and \cite{mosseretal2011} we
should identify, among all the theoretical modes, those that are most
likely to be observed, and thus contribute to the observed  period
spacing value. That would require non-adiabatic computations, which
are very time consuming and unfeasible for the large number of models
considered in this study.  Moreover, from a theoretical point of view,
the detectability of mixed modes not only depends on the models
(stellar structure and time-dependent convection) which are subject to
uncertainties, but it also strongly depends on the duration of
observations \citep{dupretetal2009} and to a lower extent on the
instrument. Therefore, following an adiabatic approach, we consider
the modes with lower inertia as those most likely to be detected.
These are also the modes contributing to significant deviations from
the uniform $\Delta P$  predicted by the asymptotic approximation
(minima in $\Delta P$, see Fig.~\ref{fig:propag}, lower panels). We
define $\langle \Delta P\rangle_{\rm th-obs}$, a theoretical
estimation of the measurable period spacing from observed oscillation
spectra, following the same procedure as in
  \cite{beddingetal2011} and based on the properties of observed
  oscillation spectra. We select oscillation modes with angular degree $\ell=0$, and 1 and with frequencies $\nu$ in the expected solar-like domain, defined as  $0.75 < \nu/\nu_{\rm max} < 1.25$, where $\nu_{\max}= M({\rm M}_\odot) R({\rm R}_\odot)^{-2} (T_{\rm eff}/5777)^{-0.5}\,3050\,\mu$Hz.   Around each pressure-dominated dipole mode we consider the values $\Delta P(n)$ involving $k$ modes with frequency lower and higher than that of the inertia minimum. Our choice ($k=2$) is based on current observations  (3 or 4 values for each radial order).  The $\Delta P(n)$'s obtained in this way are then averaged  in the solar-like frequency domain to obtain  $\langle \Delta P\rangle_{\rm th-obs}$. 

The behaviour of  $\langle \Delta P\rangle_{\rm th-obs}$ as a function of the average large frequency separation is shown in Fig.~\ref{fig:dPdnu} for models in the RGB and in the central He-B  phase with different chemical composition.  These results are in good agreement  with the recent  observational results obtained with {\it Kepler} and CoRoT, where  the clear difference between the period spacing of  RGB and He-B models allowed us to use $\langle \Delta P\rangle$ to identify the evolutionary state of red giants with comparable  $\langle \Delta\nu \rangle$ or $\nu_{\rm max}$ \citep{beddingetal2011, mosseretal2011}.

\section{Period spacing in the red-clump and secondary-clump stars: a proxy for the He-core mass}

Both the predicted and observed \citep[see][]{beddingetal2011, mosseretal2011} period spacing show significant scatter in He-B stars. This dispersion is partly due to the different mass of the stars, chemical composition, and  central helium mass fraction ($Y_{\rm C}$). \cite{mosseretal2011} identified the  high-$\Delta\nu$--low-$\Delta P$ tail as corresponding to the secondary clump, and we will show here that asymptotic and ``measurable'' values of the  period spacing contain additional information on the structure and previous evolution of the star.

\begin{figure}[ht!]
\includegraphics[scale=0.65]{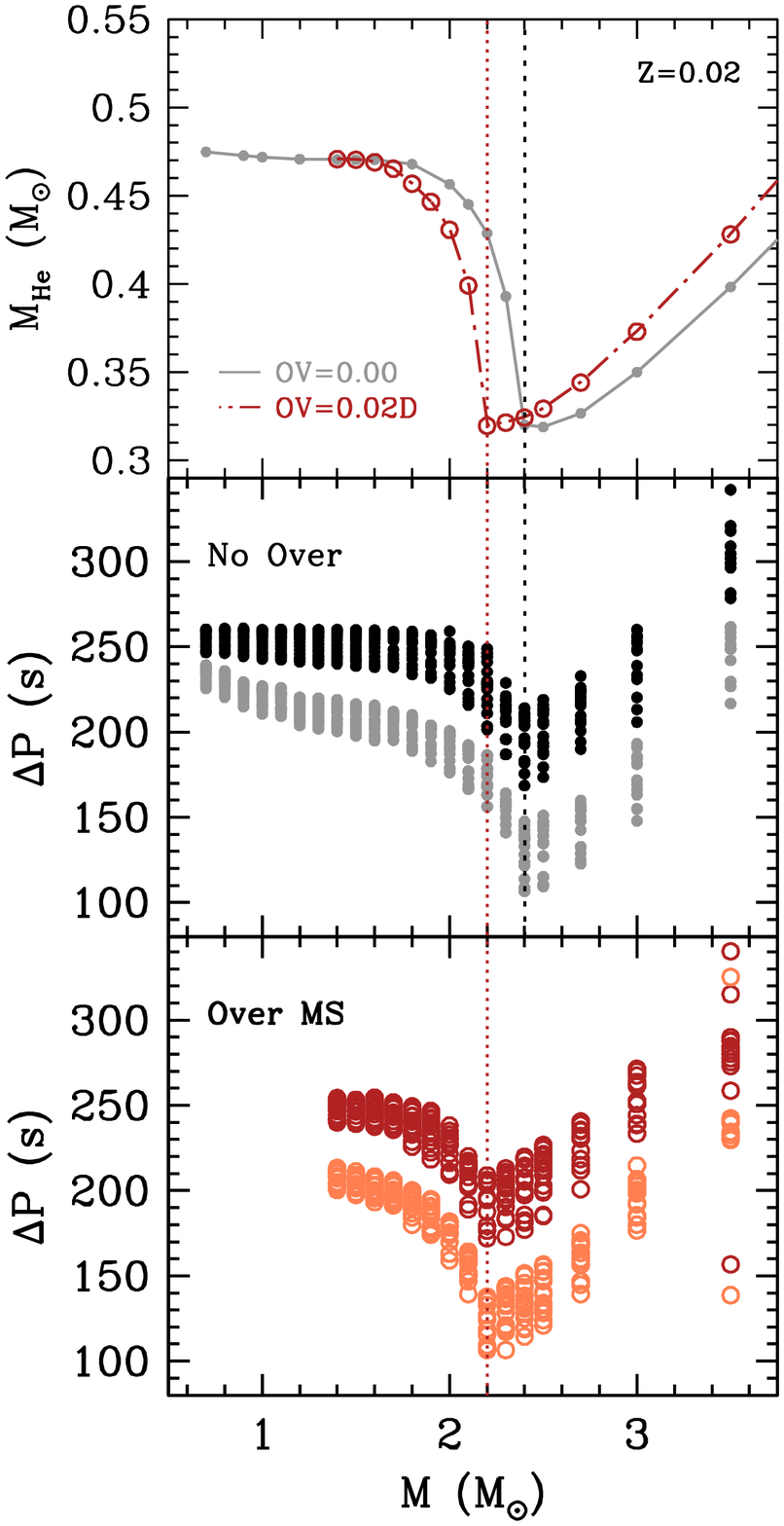}
\caption{Upper panel: He-core mass at He ignition {\it versus}
  stellar mass. Solid line and dots correspond to models  without
  overshooting during the MS phase, dashed line and open circles to
  models computed with diffusive overshooting \citep{venturaetal1998}.
 Central panel: Average period
  spacing during the He-B phase ($0.9 > Y_{\rm C}> 0.1$) {\it versus}
  stellar mass. Black dots correspond to $\langle\Delta P\rangle_a$
  and gray ones to $\langle \Delta P\rangle_{\rm th-obs}$. Lower panel:
  As central panel for models with overshooting during MS phase.
Vertical lines indicate the transition mass for no-overshooting models (small-dashed line) and MS overshooting ones (dotted line).} 
\label{fig:modes}
\end{figure}

\begin{figure*}[ht!]
\resizebox{\hsize}{!}{\includegraphics[angle=-90]{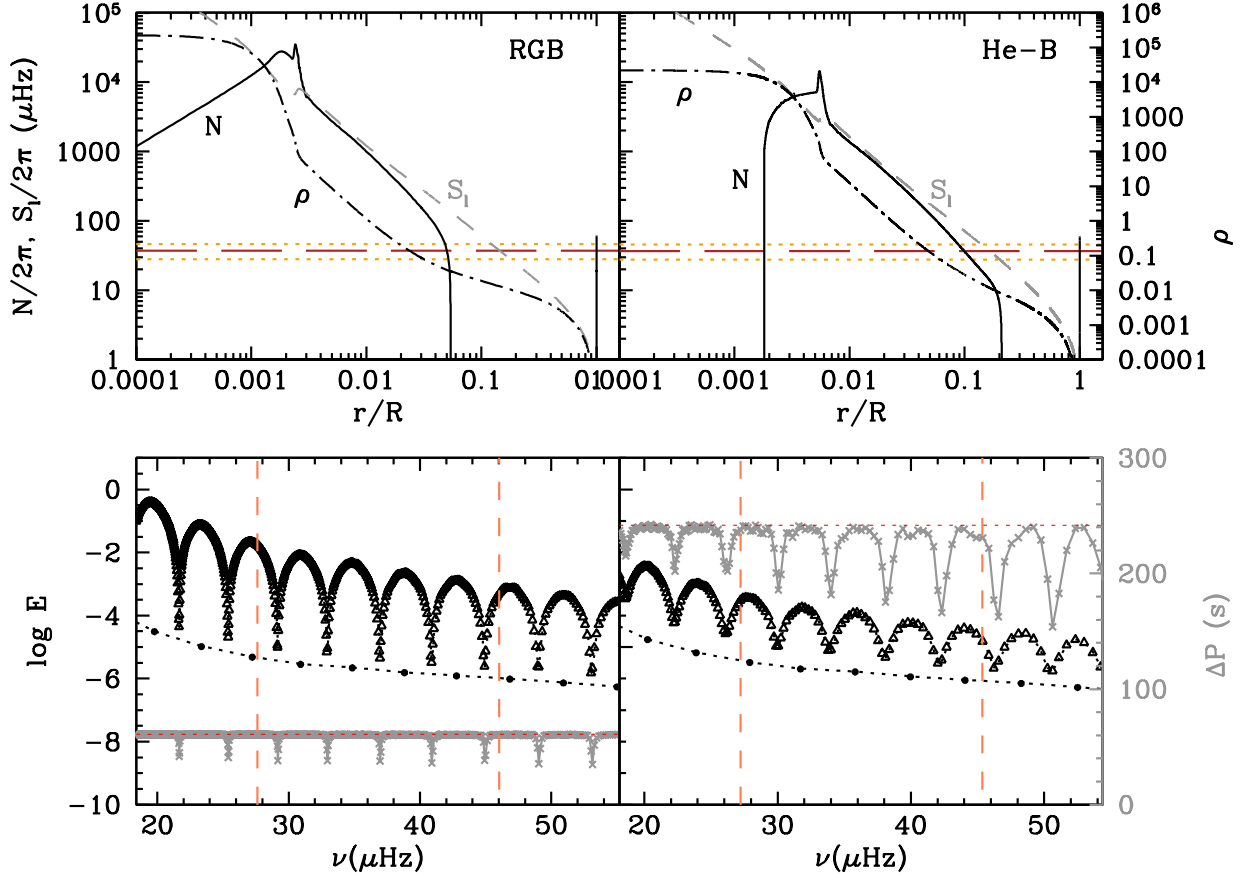}}
\caption{Upper panels:  density distribution and $\ell=1$ propagation
  diagram for a 1.5~\msun\ star with radius $\sim 12$~\rsun: in the
  ascending red giant branch (RGB, left) and in the red clump (He-B, right). 
Horizontal dotted lines denote the frequency domain of solar-like
oscillations. Solid line is \BV\ frequency, and dashed one the Lamb
frequency for $\ell=1$. Dash-dotted line represents density (scaled 
on right axis). He-B model has a small  convective core ($r_{\rm
  cc}\simeq 0.002\ R$ and $m_{\rm cc}\simeq 0.08\ M$) and its central
density is ten times smaller than for the RGB one. Due to electron
degeneracy, $N$ in the RGB model is significantly lower in the deep central layers than near the H-burning shell.
Lower panels: corresponding plots of  inertia ($E$)
{\it versus} frequency for $\ell$=0 (circles) and 1
(triangles) modes. Vertical dashed lines correspond to horizontal ones in
upper panels. Grey crosses and lines represent period separation
between consecutive dipole modes {\it versus} frequency (right
axis). Horizontal thin-dashed lines correspond to $\langle\Delta P
\rangle_{\rm a}$.}
\label{fig:propag}
\end{figure*}

  $\langle \Delta P\rangle_a$ depends mainly on the value of the $\rho_c/\langle\rho\rangle$ (density contrast), on the dimension of the
 convective core, and on the location of the H-shell which gives rise to a local  maximum of $N$. All these quantities change during the He-B phase and they do it differently depending on the stellar mass. In Fig.~\ref{fig:modes} (black dots in central panel) we plot 
$\langle \Delta P\rangle_a$ for models with masses between 0.7 and 3.5~\msun\ during the He-B phase as a function of the stellar mass. 
By comparison with the corresponding curve of the upper panel, it seems evident there is a direct correlation between 
$\langle \Delta P\rangle_a$ and the mass of the He-core. $\langle \Delta P\rangle_a$ is almost constant for low mass stars that begin 
to burn He in a degenerate core of ~0.475~\msun, and presents a minimum at the transition mass, which corresponds also to the minimum of 
He-mass core ($\sim 0.33$\msun). For higher masses, $\langle \Delta P\rangle_a$ increases with the stellar mass, such as the mass of the He-core ($\sim 0.1\,M_T$ at the end of MS).

 $\langle\Delta P\rangle_{\rm th-obs}$ as described in the
  previous section, involves  additional  information related to the
  coupling between acoustic and gravity cavities, and hence depends on
  the properties of the evanescent region which can be characterised,
  roughly speaking, by the value of the  integral $\int (\sigma
  c)^{-1} ((\sigma^2-N^2)(S_{\ell}-\sigma^2))^{1/2} dr$ in the region
  between the He-core and the envelope. The gray dots in
  Fig.~\ref{fig:modes} (central panel) represent the corresponding
  values of $\langle\Delta P\rangle_{\rm th-obs}$ for models during
  the He-B phase.   The difference between asymptotic and "observable"
  values also depends on the stellar mass for low-mass models. That
  should not be surprising since the central density of models with
  0.7~\msun\ and 1.5~\msun, their He-core mass and total radius are
  almost the same, which leads to a very different density (and
  temperature) distributions  in their envelope, and hence to a very
  different coupling.  Nevertheless,  there is still a clear
  dependence of  $\langle\Delta P\rangle_{\rm th-obs}$  on the He-core mass.

$\langle\Delta P\rangle_{\rm th-obs}$ like $\langle \Delta P\rangle_{\rm a}$ presents a minimum corresponding to $M_{tr}$,  whose  value strongly depends on the extension of mixed central region during the MS.  In Fig.~\ref{fig:over}  we plot the "observable" period spacing, for models with  and without  overshooting during the MS, as a function of the He-core mass. Two linear relationships appear between $\langle\Delta P\rangle_{\rm th-obs}$ and $M_{\rm He}$: the first for low-mass RGB models with $\Delta P$ decreasing as $M_{\rm He}$ increases due to the larger contrast density as the star evolves,  the second one for He-B models. For the latter,  both families of models follow the same relation, the only difference being the mass of the models  occupying the same location in that diagram. 

 In the framework of adiabatic calculations, that is assuming the amplitude of modes directly linked to $\sqrt E$, we use the ratio between the inertia of a mixed mode and that of the corresponding pressure-dominated one to flag that mode as detectable and thus contributing to the average period spacing value. That ratio depends on details of the stellar structure, and therefore on stellar parameters such as mass, age, and evolutionary state. In fact, we should expect to observe less mixed modes as the inertia ratio increases, which is the case for models with better trapped dipole modes. We are aware that this detectability  limit  is an arbitrary choice since the relation between inertia and amplitude is only a proxy,  nevertheless the current available non-adiabatic computations confirm that assumption \citep{dupretetal2009,grosjeanetal2013}. Moreover, we can expect more modes to be detected in longer time series, changing the value of  $\Delta P$ obtained from the mean value of individual measured period spacings, and $\langle\Delta P\rangle_{a}$ being an upper limit.

\begin{figure}[ht!]
\resizebox{\hsize}{!}{\includegraphics{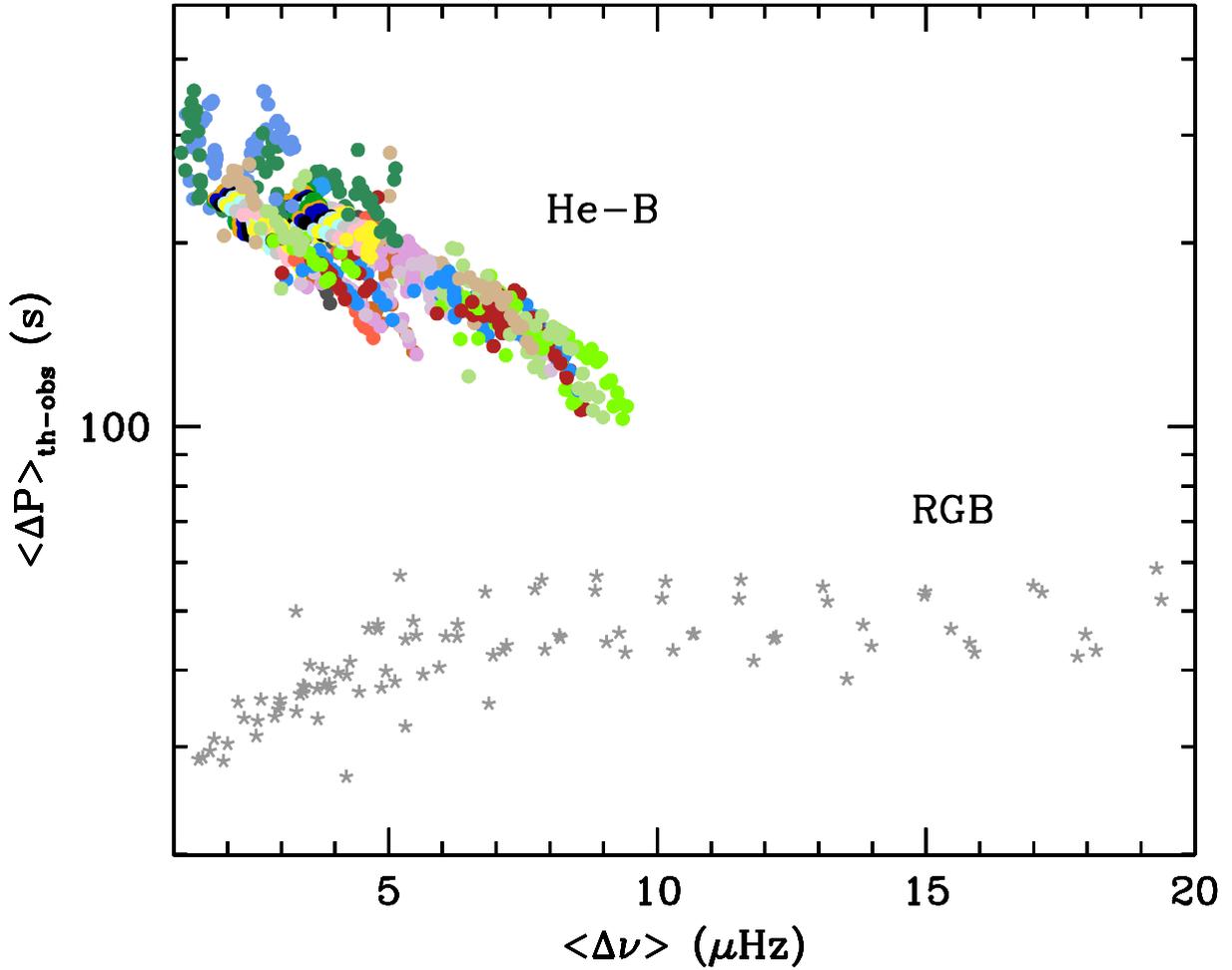}}
\caption{Theoretical ``observable'' period spacing {\it versus} large
 frequency separation of radial modes. Asterisks:  RGB
 models with masses 0.9, 1.0 1.5, 1.6 and 1.7~\msun, and chemical
 composition (Z=0.02, Y=0.278). Solid dots:  He-B models with masses
 between 0.7 and 4.0~\msun\ and chemical compositions (Z,Y):
 (0.02,0.278); (0.01,0.278);(0.02,0.25); (0.02,0.4).  Each color
   corresponds to models of different  stellar mass.}
\label{fig:dPdnu}
\end{figure}

\begin{figure}[ht!]
\resizebox{\hsize}{!}{\includegraphics{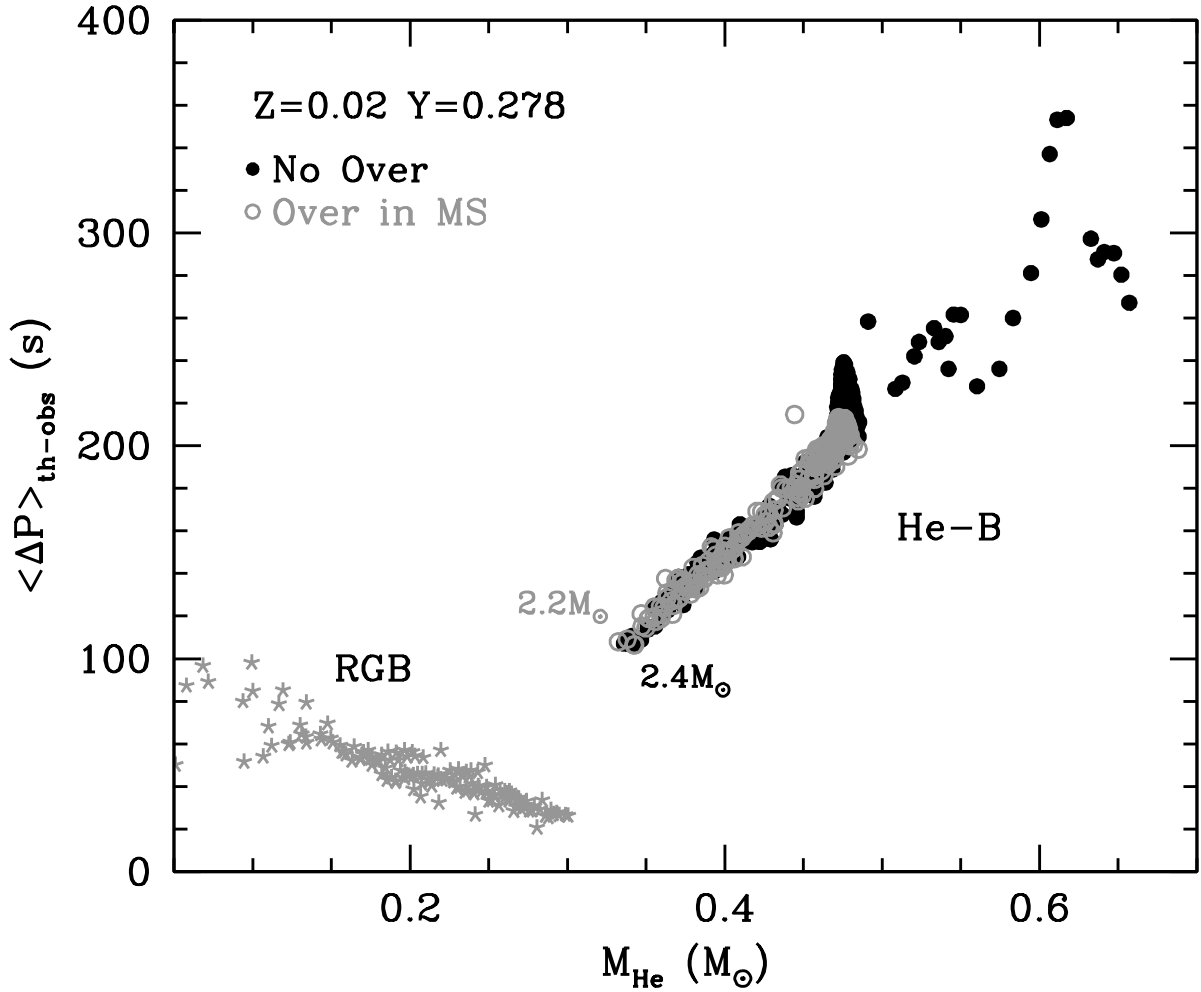}}
\caption{ Theoretical ``observable'' period spacing {\it versus} He-core mass for RGB models  (gray asterisks). Black dots: He-B  ($0.9\ge Y_{\rm C}\ge 0.1$)  models with masses between 0.7 and 4.0~\msun\ without overshooting. Open gray circles: the same for MS-overshooting models and stellar mass between 1.4 and 3.0~\msun.}
\label{fig:over}
\end{figure}

\begin{figure}[ht!]
\resizebox{\hsize}{!}{\includegraphics{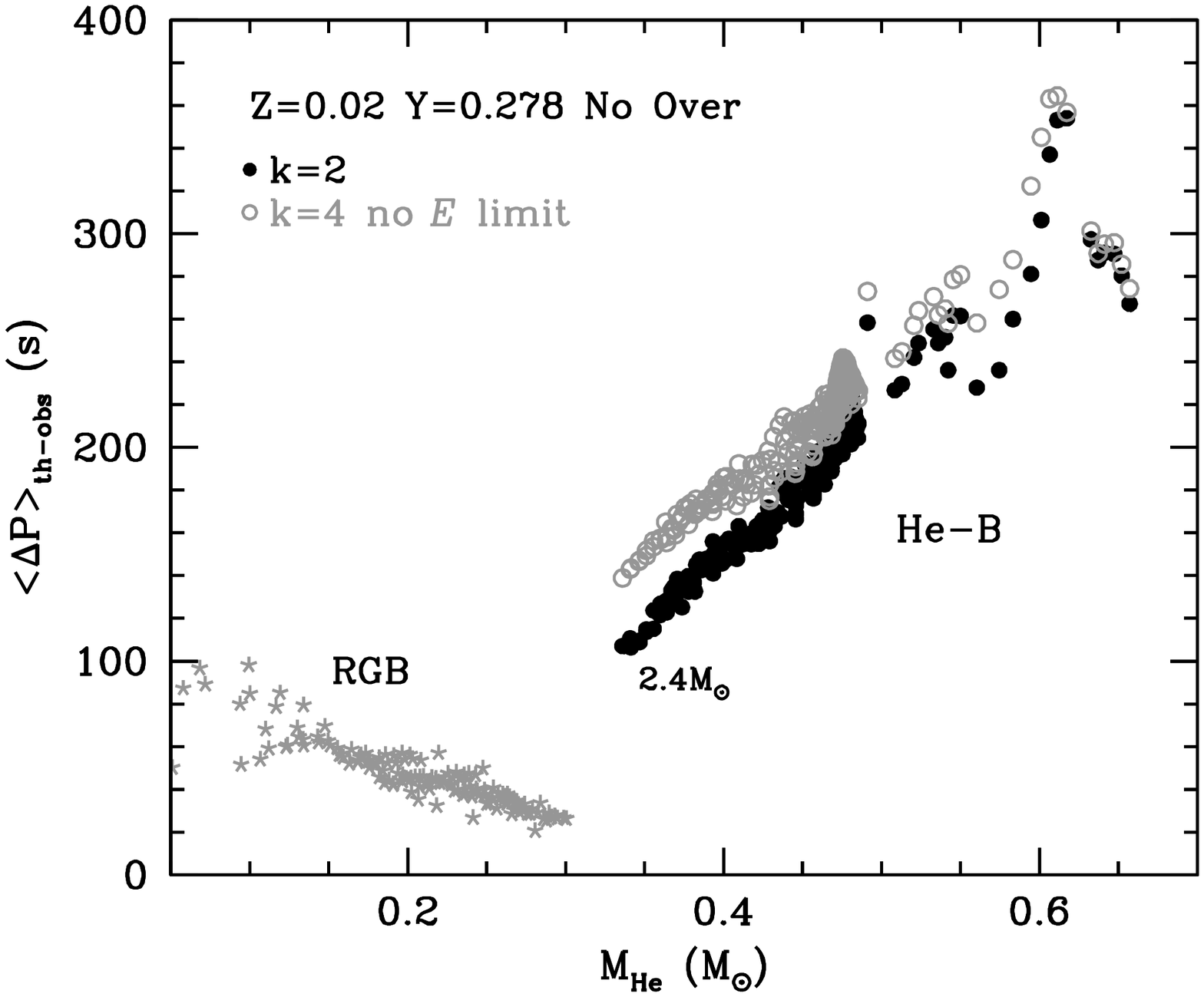}}
\caption{Theoretical ``observable'' period spacing {\it versus} He-core mass for models computed without overshooting. Black dots:  $\langle \Delta P\rangle_{\rm th-obs}$ computed with our standard method (see text); gray open circles: $\langle \Delta P\rangle_{\rm th-obs}$ computed taking 4 mixed modes at lower and higher frequencies than that of the pressure-dominated modes, and without imposing any detectability limit.}
\label{fig:test}
\end{figure}

To see how our results depend on the choice of the detectability limit and on the number of mixed-modes considered as observables, we estimated $\langle \Delta P\rangle_{\rm th-obs}$ changing these parameters. In Fig.~\ref{fig:test} we compare the results obtained for He-B models adopting  $k=2$ and a limit value of the inertia ratio (i.e., the standard procedure used in this work in Figs.~\ref{fig:modes},~\ref{fig:dPdnu}, and ~\ref{fig:over}) with the $\langle \Delta P\rangle_{\rm th-obs}$ values that results when 4 mixed modes at lower and higher frequency of each pressure-dominate mode in the solar-like frequency domain are considered, and that without imposing any detectability limit. An additional increase of $k$ does not introduce noticeable changes on the behaviour of $\langle \Delta P\rangle_{\rm th-obs}$. Note that $\langle \Delta P\rangle_{\rm th-obs}$ for lowest mass models is only slightly affected, and that the more important effect is obtained for secondary clump models. It is worth mentioning that in the case of secondary clump stars such a  large number of detected mixed modes around the pressure-dominated ones allows us to derive $\Delta P$ as a function of frequency, and then do a detailed analysis of their stellar structure.

 We stress that for a theoretical interpretation of the observations it is mandatory to use definitions of theoretical and observational indexes as close as possible. That is true for $\langle \Delta P\rangle_{\rm th-obs}$, but also for $\langle \Delta \nu\rangle$, or $\langle \delta \nu \rangle _{n\ell}$.  Even if the specific values of $\langle \Delta P\rangle_{\rm th-obs}$ may change with the number of mixed modes considered, a linear relation between $\langle \Delta P\rangle_{\rm th-obs}$ and He-core mass remains. The $k$ value must be adapted to the current observations, i.e., changing according to the duration of time series and instrument features.

Period spacing in He-B red giant stars provides a stringent test of the central mixing during the H-MS. However, to successfully exploit this possibility, spectroscopic constraints on the chemical composition are required, together with an accurate estimate of stellar mass: seismic constraints (other than period spacing) will be crucial in this respect. 

\section {Period spacing as a test of overshooting and mixing during the core-Helium burning phase}

\begin{figure}[ht!]
\resizebox{\hsize}{!}{\includegraphics[angle=-90]{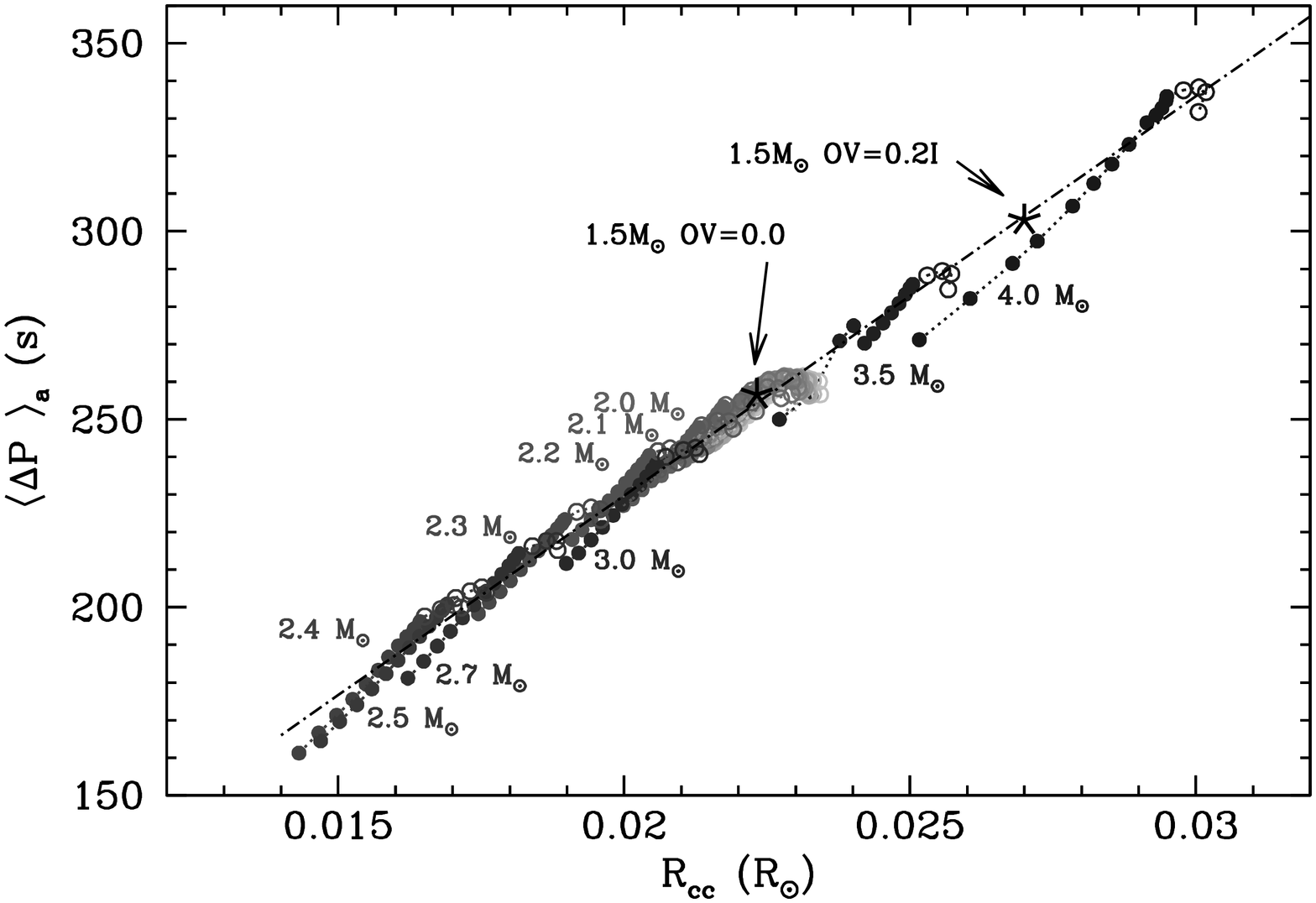}}
\caption{Asymptotic period spacing {\it versus} convective core radius for  models with masses between 0.7 and 4.0\msun in the He-B phase (solid dots : $0.9 \ge Y_{\rm C} > 0.3$, open circles $0.3\ge Y_{\rm C}\ge 0.1$).  The regression line obtained from these points: $\langle \Delta P \rangle_{\rm a}=17.35+1.06176\times 10^4\times R_{\rm cc}(R_\odot)$.  Black asterisks  correspond to two models of 1.5~\msun\ at $Y_{\rm C}=0.5$,  without and with instantaneous and adiabatic overshooting during the He-B phase.}
\label{fig:dP_rcc}
\end{figure}

 The value of  the asymptotic period spacing depends on the central distribution of $N$, and hence on the size of the convective core. Fig.~\ref{fig:dP_rcc} shows  $\langle \Delta P \rangle_{\rm a}$ as a function of convective-core radius  for models in the He-B phase, with $Y_{\rm C}$ between 0.9 and 0.1, and masses from 0.7 to 4.0\msun.  As can be seen  from  Fig.~\ref{fig:dP_rcc} the relation between $\langle \Delta P \rangle_{\rm a}$ and $R_{\rm cc}$ is, to a good approximation, linear.
 
For a given mass, the trend of $\langle \Delta P \rangle_{\rm a}$ changes when $Y_{\rm C}\simeq 0.3$ due to the much lower temperature dependence of the nuclear reaction $^{12}C(\alpha,\gamma)^{16}O$ which, at that stage, becomes the dominant energy source in the core. For massive stars  $R_{\rm cc}$ and $\langle \Delta P \rangle_{\rm a}$  decrease for models  with $Y_{\rm C} < 0.3$ because  $\rho_{\rm c}$ increases as He is exhausted. For models with $Y_{\rm c} \ga 0.3$ the mass and radius of the convective core follow a linear relation, which however disappears when $^{12}C(\alpha,\gamma)^{16}O$  takes over  the $3\alpha$ reaction: in that case $M_{\rm cc}$ increases even more quickly but $R_{\rm cc}$ decreases.

Recently \cite{jcd2011} and \cite{mosseretal2012} showed that $\langle \Delta P \rangle_a$  can be inferred from the observed period spacing of dipole mixed modes. Checking  the  procedure with  our theoretical frequencies for 1.5~\msun\ models the He-B phase, we were  able to recover the asymptotic period spacing with a precision of 2\% when we take into consideration 6--8 mixed modes for each radial order, such as in \cite{mosseretal2012} (see their Fig.~2).

\begin{figure}[ht!]
\resizebox{\hsize}{!}{\includegraphics{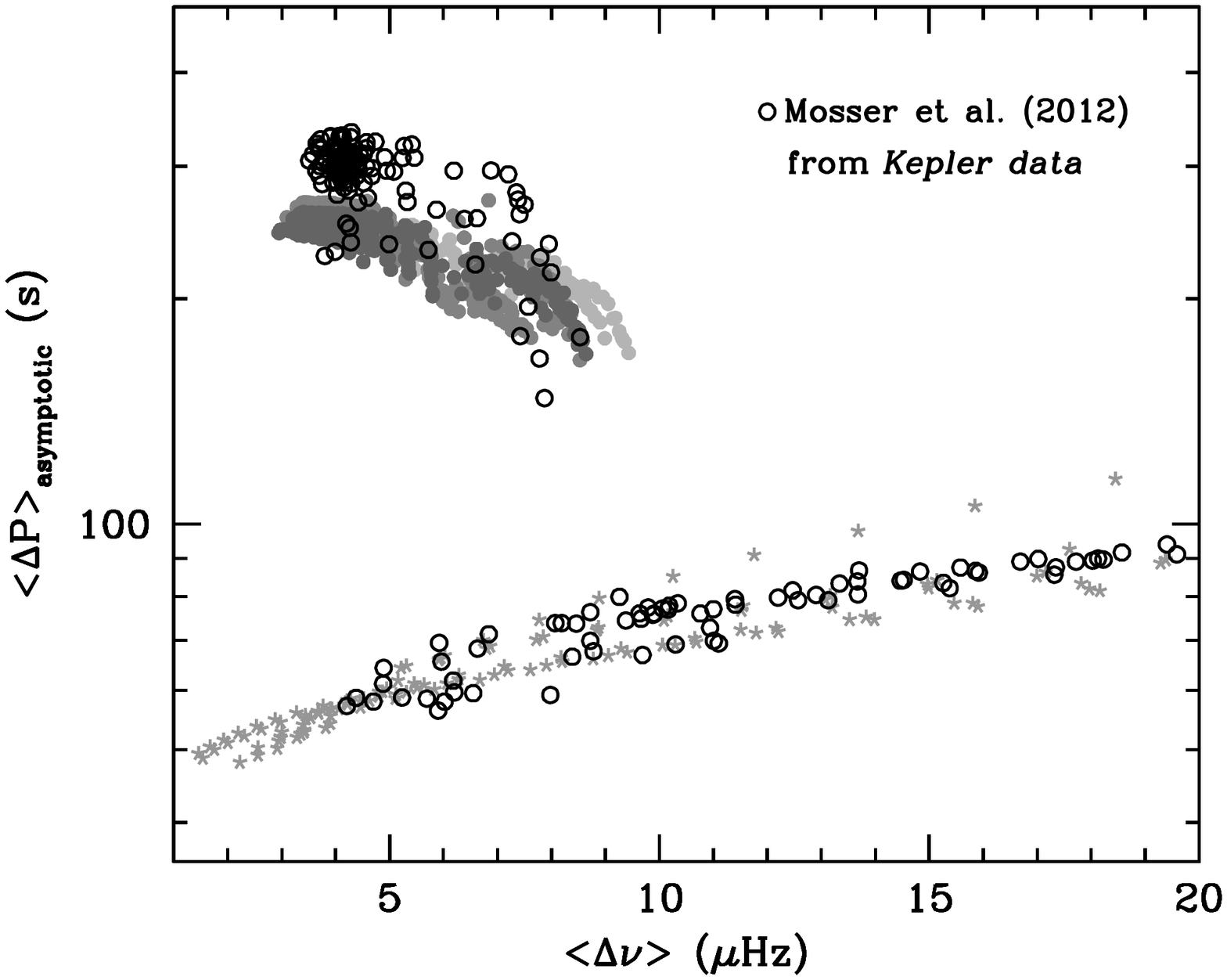}}
\caption{ Theoretical asymptotic period spacing (Eq.~\ref{eq_dp}) {\it versus} average large frequency separation of 
radial modes for models on the RGB (asterisks) with masses: 0.9, 1.0 1.5, 1.6, 1.7 and 2.0~\msun, and for He-B models 
with mases between 0.7 and 3.0~\msun\ and three different chemical compositions ($Z,Y$): (0.02,0.278) dark-gray dots, 
(0.01,0.278) middle-gray dots, (0.02,0.25) light-gray dots. 
Black open circles are the asymptotic period spacing inferred by Mosser~et~al.(2012) for a sample of red giants in the {\it Kepler} field.}
\label{fig:dPa_Mosser}
\end{figure}

A first comparison between the predictions of our models (using Eq.~\ref{eq_dp}) and the values of $\langle \Delta P \rangle_{\rm a}$ derived by  \cite{mosseretal2012} for a sample of {\it Kepler} giants is shown in Fig.~\ref{fig:dPa_Mosser}. All these models were computed without extra-mixing during the MS and during the He-B phases. It is clear that for the clump stars (low mass stars with $\Delta\nu\sim 4\,\mu$Hz)  these  theoretical models systematically underestimate $\langle \Delta P \rangle_{\rm a}$ by $\sim 20\%$.

\begin{figure}[ht!]
\resizebox{\hsize}{!}{\includegraphics[angle=-90]{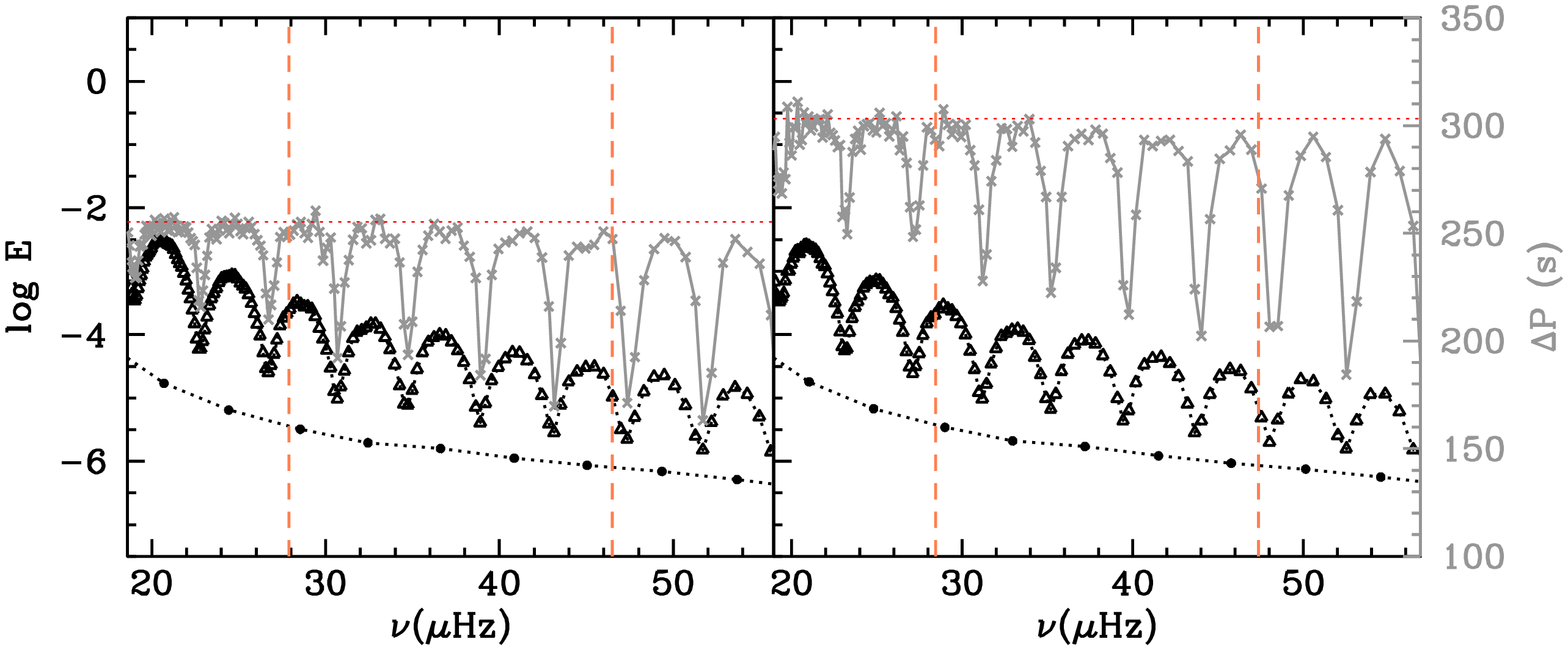}}
\caption{As lower panels in Fig.~\ref{fig:propag} for two He-B models of 1.5~\msun\ and $Y_{\rm C}=0.5$. Left:  without overshooting. Right: with overshooting during the He-B phase.}
\label{fig:over_spec}
\end{figure}

Mixing processes during the MS phase do not affect the central regions of low mass stars in He-B phase since the mass and the physical properties of the helium core are fixed by the onset of He-burning in degenerate conditions. To increase $\langle \Delta P \rangle_{\rm a}$ we should  decrease the contribution of $N$ in the innermost regions of the star, which can be easily done by extending the adiabatically stratified and chemically homogeneous core. We computed stellar models and oscillation frequencies for $M=1.5$~\msun\ with central extra-mixing during the He-B evolution described by an instantaneous and adiabatic overshooting of length $\Lambda=0.2H_p$. As expected, both the asymptotic and the minimum values of period spacing increase as the size of homogeneous core increases. In Fig.~\ref{fig:over_spec} we plot  mode inertia and period spacing {\it versus} frequency for two models, computed with and without overshooting,  of 1.5~\msun\ at $Y_{\rm C}=0.5$, and in Fig.~\ref{fig:dP_rcc} we highlight with asterisks  the values of $\langle \Delta P \rangle_{\rm a}$ for these  models: without overshooting  $\langle \Delta P \rangle_{\rm a}=255$s and with overshooting $\langle \Delta P \rangle_{\rm a}=305$s. Note that the last values ($\langle \Delta P \rangle_{\rm a}$ and $R_{\rm CC}$) are similar to those of a 4.0~\msun\ without overshooting.  The exact $\langle \Delta P \rangle_{\rm a}$  value  depends on the extension of the extra-mixed region, but also on the details of the mechanism at the origin of different $N$ profiles: a mechanical overshooting with adiabatic temperature gradient in the extra-mixed region, or an induced overshooting with a semi-convective layer outside the convective core. The latter will provide, for the same extension of the adiabatic region, a smaller period spacing than the former.  The behaviour shown in Fig.~\ref{fig:dP_rcc}  for $Z=0.02$ is also found for different metallicities. The only difference is the minimum values of $\langle \Delta P\rangle_{\rm a}$ and $R_{\rm cc}$ that slightly depend on $Z$.
Similar results were found using the MESA code \citep{MESA}, considering models with and without overshooting during the He-B phase.

\section{Conclusions}

From the computation of stellar models with masses from 0.7 to 4.0~\msun and from the study of their adiabatic oscillation spectra, we find a good agreement between theoretical and observational behaviour of period spacing. We also show the potential of dipole-mode period spacing  in RGs to constrain the extension of central extra-mixing during the H- and He-central burning phases:
\begin{itemize}
\item The ``observed''  period spacing of stars with masses close to the transition value (1.8 - 2.4~\msun\ depending on  chemical composition and mixing)  presents an almost linear relation with the He-core mass and, therefore,  contains information on the extension of the convective core during the H-MS. These stars define  the so-called ``secondary clump'', and the value of their mass has been identified as the best index of the mixing processes during the MS in stellar clusters\citep{girardi1999}. Currently, seismology allows us to derive mass and secondary clump membership   for a huge number of field RGs. Nevertheless, a fruitful exploitation of the enormous potential of seismic properties,  allowing  the comparison of observational results with predictions from synthetic stellar populations models,  would also require  information on chemical composition. So,  a theoretical $\langle\Delta P\rangle_{\rm th-obs}$ as function of the total mass, initial chemical composition ($Z,Y$) and evolutionary state ($Y_{\rm C}$ or $M_{\rm He}$) will provide the expected  $\langle\Delta P\rangle_{\rm th-obs}$ distribution  for synthetic stellar populations. Its comparison with observed distributions will shed some light on the mixing processes during the H-MS.
It is also worth recalling that stars in the secondary clump spend long time in the He-B phase, and that their ages depend mainly on the MS extra-mixing (38\% larger for stars with MS overshooting) but are chemical composition independent \citep{sweigartetal1990}. 

\item The seismic properties of stars in the red clump  seem  independent of  core extra-mixing during MS, as low-mass stars reach the same kind of structure at He ignition.  We have shown, however,  that  a linear relation exists between the asymptotic period spacing of dipole modes and the extension of the convective core in the He-B phase.  Non extra-mixing He-B models predict $\langle\Delta P\rangle_a \sim\,$20\% smaller than values inferred from observations.  If $\langle \Delta P\rangle_{\rm a}$ can be reliably derived from oscillation spectra, as suggested by \cite{mosseretal2011} and \cite{jcd2011}, the basic seismic properties of oscillation spectra for red clump stars can be used to constrain the  extension and properties of core extra-mixing during the He-B phase.
\end{itemize}

When individual frequencies of dipole mixed modes will be available, a more direct comparison between models and observations shall be carried out, and will provide information not only on the extension of the central mixed region, but also on the detailed properties of the chemical composition gradient inside the star.

\end{document}